\documentclass[a4paper,12pt]{article}
\usepackage{physics}
\usepackage{graphicx}
\usepackage{epstopdf}
\usepackage{wrapfig}
\usepackage{amsmath}
\usepackage{amsfonts}
\usepackage{amssymb}
\usepackage{color}
\usepackage{subcaption}
\usepackage{dsfont}
\usepackage{mathrsfs}
\usepackage{dcolumn}
\usepackage{bm}
\usepackage{hyperref}
\usepackage{verbatim}
\usepackage{dsfont}
\usepackage{float}
\usepackage{lipsum}
\usepackage{cite}
\usepackage[title]{appendix}
\usepackage{pgfplots}
\usepackage{amsmath}

\usepackage{tikz}

\begin{document}
\title{\textbf{Relativistic quantum Otto engine driven by the circular Unruh effect}}
\author{
{\bf {\normalsize Rudra Prosad Sarkar}\thanks{rudraprosad.sarkar@bose.res.in}},
{\bf {\normalsize Arnab Mukherjee}\thanks{arnab.mukherjee@bose.res.in}},
{\bf {\normalsize Sunandan Gangopadhyay}\thanks{ sunandan.gangopadhyay@gmail.com}}\\
{\normalsize Department of Astrophysics and High Energy Physics},\\
{\normalsize S.N. Bose National Centre for Basic Sciences},\\
{\normalsize JD Block, Sector III, Salt Lake, Kolkata 700106, India}\\
}
\maketitle
\begin{abstract}
\noindent In this work, we present a new framework for a relativistic quantum analouge of the classical Otto engine. Considering a single qubit as the working substance, we analyse its interaction with a massless quantum scalar field while undergoing two half-circular rotations at ultra-relativistic velocities. The quantum vacuum serves as a thermal bath through the Unruh effect induced due to the acceleration from the circular motions. We observe that the response function of the qubit gets significantly modified by the presence of the qubit's trajectory. Analysing the transition probability behaviour, we find that in the high-acceleration regime, it asymptotically approaches a constant value, determined solely by the properties of the correlation function. Furthermore, our results emphasize the crucial role of the circular trajectory in determining the engine's work output. In particular, the extracted work increases with detector acceleration and approaches an asymptotic limit in the high-acceleration regime. Notably, the efficiency of this model remains unaffected by the circular motion and is consistent with previously studied models.
\end{abstract}
\section{Introduction}\label{sec:Intro}
The quest to extend classical thermodynamic cycles into the quantum realm began with the pioneering works \cite{kieu2004second, kieu2006quantum, quan2007quantum, quan2009quantum, maruyama2009colloquium}, where two-level quantum systems were introduced as the working substance. This groundbreaking approach has since fuelled a plethora of investigations \cite{rezek2006irreversible, wang2009performance, abe2011similarity, thomas2011coupled, kosloff2017quantum, agarwal2013quantum, rossnagel2013nano, azimi2014quantum, zhang2014quantum, ivanchenko2015quantum, de2019efficiency, camati2019coherence, del2022quantum}, uncovering profound insights and remarkable results that continue to reshape our understanding of quantum thermodynamics.

Since the pioneering works of Fulling, Davies, and Unruh \cite{fulling1973nonuniqueness, davies1975scalar,unruh1976notes}, it has been firmly established that a uniformly accelerating observer in Minkowski spacetime perceives the Minkowski vacuum of a quantum field as a thermal bath. This phenomenon, characterized by excitations and de-excitations akin to those induced by a thermal state, is governed by the Unruh temperature $T_{U}=\hbar a/(2\pi k_{B} c)$, which depends on the observer’s proper acceleration $a$ \cite{birrell1984quantum,crispino2008unruh}. In 2018, Arias et al. \cite{arias2018unruh} pioneered the application of the Unruh effect within the framework of quantum thermodynamics, a field that has gained significant research interest in recent years \cite{gemmer2009quantum, kosloff2013quantum,alicki2018introduction,deffner2019quantum}. Their work extended the study of quantum Otto engines to the relativistic regime, exploring the impact of acceleration on thermodynamic processes. In recent years, several important studies have been done related to relativistic studies of quantum Otto engines \cite{gray2018scalar,xu2020unruh,mukherjee2022unruh,kollas2024exactly,yoshimura2024relativistic}. Some seminal works have already established the relationship between thermodynamics, relativistic quantum mechanics and black hole physics \cite{bekenstein2020black, bardeen1973four, hawking1975particle, hawking1976black}, which makes the investigation of relativistic quantum analogous models of classical heat engines a very important topic.

From the previous studies related to the Unruh Quantum Otto Engine (UQOE) \cite{arias2018unruh,gray2018scalar,xu2020unruh,mukherjee2022unruh,kollas2024exactly,yoshimura2024relativistic}, it has been observed that using the quantum fields as the hot and cold thermal baths and getting the temperature of the order of $\sim 1 K$, the acceleration of the detector needs to be of the order of $\sim 10^{20}\,m/s^2$. Such extreme accelerations are generally challenging to achieve, particularly in the case of linear acceleration. Consequently, this poses a significant obstacle to the practical implementation of UQOE models.

To address this challenge, we draw insights from the seminal works \cite{bell1983electrons, bell1987unruh}. For circular motion, such as that of electrons in the storage ring at LEP, it was observed that centripetal accelerations of the order of \( \sim 2.9 \times 10^{23} \) m/s\(^2\) can be achieved. This corresponds to an Unruh temperature of approximately $1200$ K. Such a temperature is expected to produce observable effects, and it was indeed argued that this Unruh temperature could lead to an increased population of electrons in the upper energy level beyond the expected equilibrium distribution. For oxygen and fluorine atoms, the circular Unruh effect could possibly be verified \cite{rad2012test}.

Given the feasibility of achieving the high accelerations required for experimental verification of the Unruh effect in circular motion \cite{bell1983electrons}, this approach is widely utilized for investigating the entanglement dynamics of accelerated atoms interacting with quantum fields \cite{she2019entanglement,zhang2020entanglement}. 

Recent studies \cite{bunney2023circular,bunney2024ambient,zhou2025significant,katayama2025circular} suggest that the circular Unruh effect can have a substantial impact even at relatively small centripetal accelerations. Specifically, in Ref. \cite{zhou2025significant} it reports that the excitation rate of centripetally accelerated atoms can exceed that of linearly accelerated atoms with the same acceleration magnitude by a factor of approximately $10^{272.878}$. This remarkable enhancement underscores the potential significance of circular motion in the study of acceleration-induced quantum effects. Recently, an experimentally viable scheme have been proposed in Ref. \cite{katayama2025circular} to observe the Fulling–Davies–Unruh (FDU) effect in circular motion by employing a novel detector architecture based on coupled annular Josephson junctions.

Motivated by the above studies, in this paper we propose a novel relativistic quantum analogue of the classical Otto engine to investigate the effects of the circular Unruh effect and extract quantum work, which may be measurable in future experimental setups. To achieve this, we adopt the fundamental framework of the Unruh Quantum Otto Engine (UQOE) \cite{arias2018unruh}, where a qubit undergoes a closed one-dimensional trajectory and interacts with a massless quantum scalar field in two distinct regions. Typically, in these regions, the qubit experiences accelerations of different magnitudes, allowing the quantum fields to serve as thermal reservoirs. In our model, to incorporate the effects of the circular Unruh effect, we consider that in the regions where the qubit interacts with the massless quantum scalar field, it follows circular motion along two distinct arcs with radii \(R_1\) and \(R_2\), respectively. These radii correspond to the hot and cold thermal reservoirs of the engine.
 
The paper is structured as follows: In Section \ref{sec:UQOE}, we review the fundamental principles of the Unruh Quantum Otto Engine (UQOE). Section \ref{sec:CUQOE} introduces a novel relativistic extension of the quantum Otto cycle. In Section \ref{sec:atom-field}, we analyze the interaction between a qubit and the quantum vacuum, highlighting modifications to the correlation function due to circular motion. Section \ref{sec:resfunct} focuses on evaluating the response function, a crucial quantity for determining the transition probability. In Section \ref{sec:analysis}, we perform a thermodynamic analysis of the quantum heat engine. Section \ref{sec:result} presents a detailed investigation of the effects of circular motion on transition probability, work output, and efficiency. Finally, conclusions are drawn in Section \ref{sec:conclu}.
\section{Unruh quantum Otto engine}\label{sec:UQOE} 
In quantum thermodynamics, quantum heat engines and batteries have emerged as key topics due to their potential for practical applications \cite{bhattacharjee2021quantum}. In this context, we begin by reviewing the fundamental principles of the Unruh quantum Otto engine \cite{arias2018unruh, gray2018scalar}. 

To construct such an engine, the conventional thermal reservoir is replaced by a quantum scalar field, while the working system---modeled as a detector (a qubit)---undergoes acceleration and interacts with the field over a finite time interval. Since the Unruh effect depends on the detector's motion as it interacts with the quantum vacuum, a complete characterization of its kinematic trajectory over the entire cycle is essential.  

We consider a two-level qubit system with ground state $\vert g\rangle$ and excited state $\vert e\rangle$, where the energy of the excited state is $\mathcal{E}_1$, leading to the qubit Hamiltonian  
\[
H = \mathcal{E}_1\ketbra{e}{e}.
\]  
The Otto cycle is assumed to begin with the system in an initial mixed state  
\[
\rho_0 = p\ketbra{e}{e} + (1-p)\ketbra{g}{g}.
\]  
In addition to following a thermodynamic closed cycle, the detector undergoes a kinematically closed trajectory in one-dimensional motion through Minkowski spacetime, consisting of the following four steps.  

\begin{enumerate}
    \item \textbf{Uniform Motion}: The detector moves at a constant velocity $v$ for a time duration $\mathcal{T}$.
    \item \textbf{Accelerated Motion (Hot Bath Interaction)}: The detector undergoes constant acceleration $a_H$ for a time $\mathcal{T}_2$, transitioning from velocity $v$ to $-v$.
    \item \textbf{Uniform Motion}: The detector moves at a constant velocity $-v$ for a time duration $\mathcal{T}$.
    \item \textbf{Accelerated Motion (Cold Bath Interaction)}: The detector undergoes constant acceleration $a_C$ for a time $\mathcal{T}_1$, transitioning from velocity $-v$ back to $v$.
\end{enumerate}

The interaction times $\mathcal{T}_1$ and $\mathcal{T}_2$ are determined by the values of $v$, $a_H$, and $a_C$. To ensure that the vacuum behaves as a thermal reservoir, step 2 corresponds to a hot bath, while step 4 corresponds to a cold bath, the condition $a_H > a_C$ must be satisfied.  

In analogy with the thermodynamic cycle of the engine, here we have the following steps.  

\begin{itemize}
    \item \textbf{Adiabatic Stages (Steps 1 \& 3)}: The system undergoes an adiabatic expansion and contraction of its energy gap, from $\mathcal{E}_1$ to $\mathcal{E}_2$ and back to $\mathcal{E}_1$, respectively. No heat is exchanged with the environment during these steps, but work is performed.
    \item \textbf{Thermalization Stages (Steps 2 \& 4)}: The detector interacts with the quantum field for durations $\mathcal{T}_2$ and $\mathcal{T}_1$, leading to heat exchange. The system state evolves as follows.
    After step 2, the state changes to
        \begin{eqnarray}
        \rho = (p+\delta p_H)\ketbra{e}{e} + (1-p-\delta p_H)\ketbra{g}{g},\nonumber
        \end{eqnarray}
        where $\delta p_H$ represents the transition probability induced by interaction with the hot bath.\\
        After step 4, the state further evolves to
        \begin{eqnarray}
        \rho_f = (p+\delta p_H+\delta p_C)\ketbra{e}{e} + (1-p-\delta p_H-\delta p_C)\ketbra{g}{g},\nonumber
        \end{eqnarray}
        where $\delta p_C$ accounts for the transition probability due to interaction with the cold bath. No work is performed in these steps.
\end{itemize}

For the cycle to be completed, the condition  
\[
\delta p_H + \delta p_C = 0
\]  
must hold, ensuring the system returns to its initial state after one full operation of the engine.

\section{Circular Unruh quantum Otto engine}\label{sec:CUQOE}
The influence of the linear accelerated motion of a detector on the performance of a quantum heat engine has been extensively analysed within the framework of the quantum Otto cycle \cite{arias2018unruh,gray2018scalar,xu2020unruh,mukherjee2022unruh,kollas2024exactly,yoshimura2024relativistic}. From previous studies, it is revealed that while the accelerated motion of the detector leads to a modification in the output work of the engine, it does not affect its efficiency. This suggests that the fundamental thermodynamic efficiency of the quantum Otto engine remains robust against the influence of linear acceleration, even as the work output undergoes significant changes. 

Our objective in this study is to explore how the circular motion of a detector influences the efficiency and work output of a quantum heat engine. A striking aspect of the circular Unruh effect is that it can manifest significantly even at relatively small centripetal accelerations, challenging the conventional belief that a strong Unruh effect requires extremely high acceleration \cite{zhou2025significant}.  This intriguing property not only deepens our understanding of relativistic quantum thermodynamics but also sheds new light on the experimental feasibility of detecting quantum work extraction, offering potential pathways for future advancements in the field.

\begin{figure}[h!]
    \centering
    \includegraphics[width=0.65\textwidth]{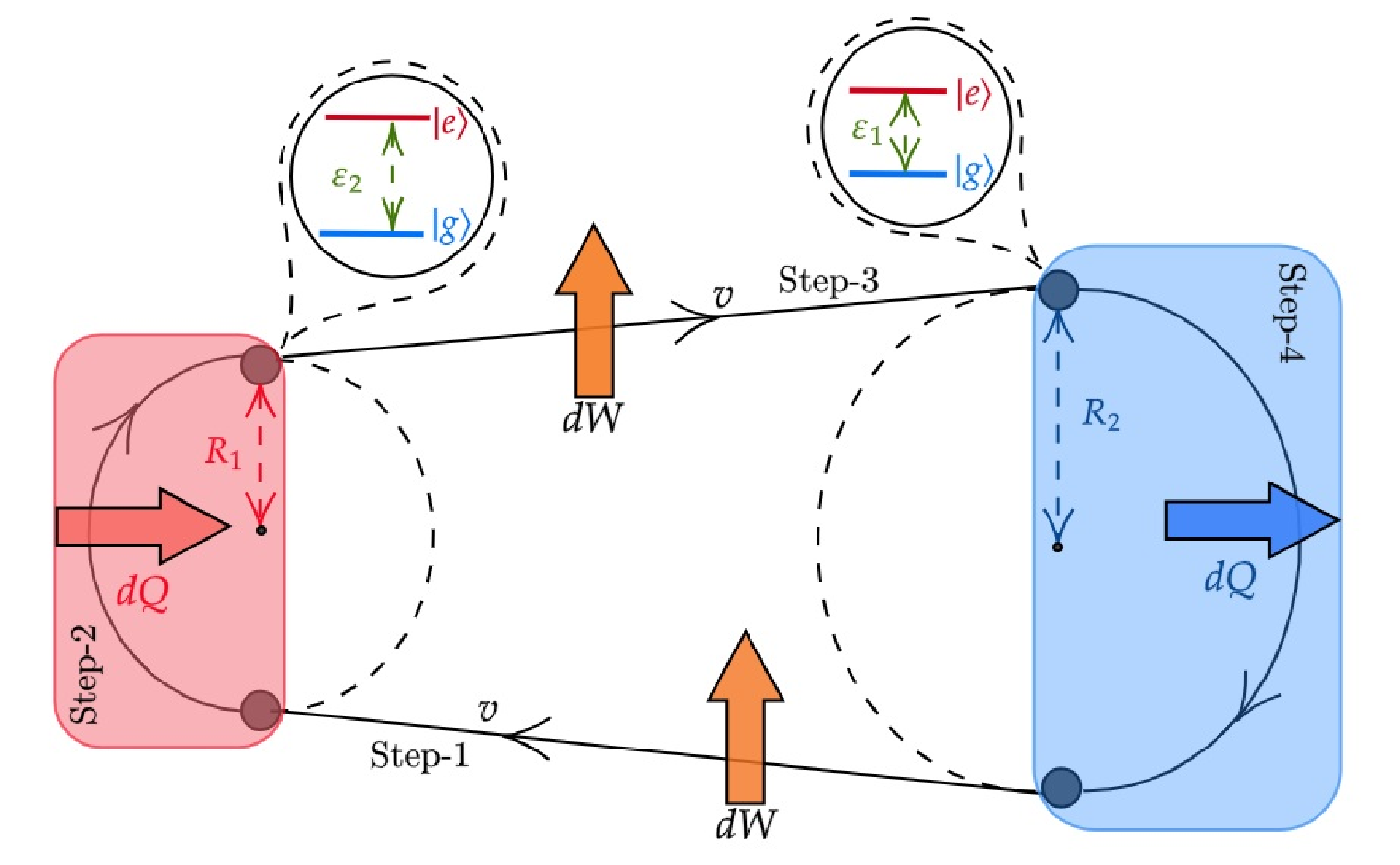}
    \caption{Thermodynamic cycle of UQOE along a circular motion.}
    \label{fig:CUQOE}
\end{figure}

\noindent From Fig. \ref{fig:CUQOE}, it is observed that during the motion along a closed trajectory through Minkowski spacetime, it is following a thermodynamic closed cycle. The entire motion of the detector can be separated into 4 different stages. 

\begin{enumerate}
    \item \textbf{Uniform Motion and Adiabatic Expansion (Step-1)}: The detector moves at a constant velocity $v$ for a time duration $\mathcal{T}$. During this time the system undergoes an adiabatic expansion of its energy gap, from $\mathcal{E}_1$ to $\mathcal{E}_2$. No heat is exchanged with the environment during these steps, but some external work is performed on the system.
    \item \textbf{Circular Motion (Hot Bath Interaction)}: The detector undergoes a circular motion along the perimeter of a circle of radius $R_1$ for a time $\mathcal{T}_1$. In this step, centripetal acceleration will be $a_{c_{Hot}}=\frac{\gamma^2 v^2}{R_1}$ where $\gamma=\frac{1}{\sqrt{1-v^2}}$ and $\mathcal{T}_1=\frac{\pi R_1}{\gamma v}$.
    \item \textbf{Uniform Motion and Adiabatic Expansion (Step-3)}: The detector moves at a constant velocity $v$ in the opposite direction of step-1. During this time the system undergoes an adiabatic compression of its energy gap, from $\mathcal{E}_2$ to $\mathcal{E}_1$. Here also no heat is exchanged with the environment, but the system will perform some external work on the environment. This step is often called the \textit{power stroke}.
    \item \textbf{Circular Motion (Cold Bath Interaction)}: The detector undergoes a circular motion along the perimeter of a circle of radius $R_2$ for a time $\mathcal{T}_2$. In this step, centripetal acceleration will be $a_{c_{Cold}}=\frac{\gamma^2 v^2}{R_2}$ and $\mathcal{T}_2=\frac{\pi R_2}{\gamma v}$.
\end{enumerate}

\noindent From the above Figure, it is observed that $R_1<R_2$, therefore $a_{c_{Hot}}>a_{c_{Hot}}$. Apart from this, we also assume that the centres of the two circles are far apart from each other compared to their radius. Thus, in each of the two circular paths, the detector completes the half rotation of the circles.

\noindent In a circular trajectory in $3+1$ dimensions, we have
\begin{eqnarray}\label{trajectory}
    t(\tau)=\gamma \tau,\,\,\,x(\tau)=R\cos{\big(\Omega\gamma \tau\big)},\,\,\, y(\tau)=R\sin{\big(\Omega\gamma \tau\big)},\,\,\, z(\tau)=0
\end{eqnarray}
where $R$ is the radius of the circle and 
\begin{eqnarray}
\Omega=\frac{d\theta}{d\tau}=\frac{v}{R}~.
\end{eqnarray}

\noindent To account for the fact that the UD detector does not complete full rotation along the perimeter, we adopt as a regulator $ \varpi_{\mathcal{T}}(\tau)$ a Lorentzian profile to work as switching function, that is,
\begin{eqnarray}
\varpi_{\mathcal{T}}(\tau)=\frac{(\mathcal{T} / 2)^2}{\tau^2+(\mathcal{T} / 2)^2}~.
\end{eqnarray}

\section{Interaction of the field vacuum with the qubit}\label{sec:atom-field}

The Unruh-Dewitt system total Hamiltonian is made up of three parts. This reads
\begin{eqnarray}
\mathcal{H}=\mathcal{H}_{qubit}+\mathcal{H}_{field}+\mathcal{H}_{int}
\end{eqnarray}
where
\begin{eqnarray}
\mathcal{H}_{qubit}&=&\mathcal{E}\ketbra{e}{e}\\
\mathcal{H}_{field}&=&\int \text{d}^{3}k\,\mathcal{E}(k)\,{b}^{\dagger}(\mathbf{k})b(\mathbf{k})\\
\mathcal{H}_{int}&=&\lambda\,m(\tau)\,\phi[x(\tau)].
\end{eqnarray}
Here $b(\mathbf{k})$ and $b^{\dagger}(\mathbf{k})$ are the bosonic operators of the scalar field \cite{peskin2018introduction}, $\lambda$ is a weak qubit field coupling constant and $m(\tau)$ is the monopole operator of the qubit.

\begin{eqnarray}
m(\tau)=e^{\frac{i\mathcal{E}\tau}{\hbar}}\ketbra{e}{g}+e^{-\frac{i\mathcal{E}\tau}{\hbar}}\ketbra{g}{e}=\begin{pmatrix}
0 & e^{\frac{i\mathcal{E}\tau}{\hbar}} \\
e^{-\frac{i\mathcal{E}\tau}{\hbar}} & 0
\end{pmatrix}~.
\end{eqnarray}
As we can see, in this model we consider that the qubit acts as a point particle and the total Hamiltonian is written in the interaction picture so that we can use the free mode expansion of the massless scalar field.

\subsection{Evolution of the detector} Let us assume that initially the density matrix of the UD detector is given by

\begin{eqnarray}
\rho_0=p\ketbra{e}{e}+(1-p)\ketbra{g}{g}=\begin{pmatrix}
p & 0 \\
0 & 1-p
\end{pmatrix}~.
\end{eqnarray}

The initial density matrix of of the scalar field in an inertial frame having the vacuum state $|0\rangle$ is given by

\begin{eqnarray}
    \rho_{field}=\ketbra{0}{0}~.
\end{eqnarray}

After interaction time $\mathcal{T}$, the density function becomes 

\begin{eqnarray}
\rho_\mathcal{T}=\begin{pmatrix}
p+\delta p_\mathcal{T} & 0 \\
0 & 1-p-\delta p_\mathcal{T}
\end{pmatrix}~.
\end{eqnarray}

Here the transition probability arising due to quantum vacuum fluctuations between the levels of the detector is given by $\delta p_\mathcal{T}$. Thus, we can write the transition probability as \cite{arias2018unruh}

\begin{eqnarray}
    \delta p_\mathcal{T}=\frac{\lambda^2}{\hbar^2}\int_{-\mathcal{T}}^\mathcal{T}d\tau \int_{-\mathcal{T}}^\mathcal{T}d\tau'\Big[(1-p)e^{-\frac{i\mathcal{E} \Delta\tau}{\hbar}}-pe^{\frac{i\mathcal{E} \Delta\tau}{\hbar}}\Big]\mathcal{G}^+(\tau, \tau')~.
\end{eqnarray}
Introducing the switching function $\varpi_{\mathcal{T}}(\tau)$ and taking the limit $\mathcal{T}\rightarrow\infty$, the above equation takes the form
\begin{eqnarray}
    \delta p_\mathcal{T}=\frac{\lambda^2}{\hbar^2}\int_{-\infty}^\infty d\tau \int_{-\infty}^\infty d\tau'\varpi_{\mathcal{T}}(\tau)\varpi_{\mathcal{T}}(\tau')\Big[(1-p)e^{-\frac{i\mathcal{E} \Delta\tau}{\hbar}}-pe^{\frac{i\mathcal{E} \Delta\tau}{\hbar}}\Big]\mathcal{G}^+(\tau, \tau')\,.
\end{eqnarray}

\noindent Here, $\mathcal{G}^+(\tau, \tau')$ is called the positive frequency Wightman function \cite{birrell1984quantum}
\begin{eqnarray}
    \mathcal{G}^+(\tau, \tau')=\langle 0|\phi[x(\tau)]\phi[x(\tau')]|0\rangle\,.
\end{eqnarray}
Defining the response function as 
\begin{eqnarray}
    \mathcal{F}(\mathcal{E}, \mathcal{T})=\frac{1}{\hbar^2}\int_{-\infty}^\infty d\tau \int_{-\infty}^\infty d\tau'\varpi_{\mathcal{T}}(\tau)\varpi_{\mathcal{T}}(\tau')\mathcal{G}^+(\tau, \tau')e^{-\frac{i\mathcal{E} \Delta\tau}{\hbar}}
\end{eqnarray}

we can write
\begin{eqnarray}\label{deltap}
    \delta p_{\mathcal{T}}=\lambda^2\Big[(1-p)\mathcal{F}(\mathcal{E}, \mathcal{T})-p\mathcal{F}(-\mathcal{E}, \mathcal{T})\Big]~.
\end{eqnarray}

\section{Evaluation of the response function}\label{sec:resfunct}
In circular motion the Wightman function only depends on the difference in time \cite{yoshimura2023quantum}.
The Wightman function in flat space-time is given by
\begin{eqnarray}
    \mathcal{G}(\tau, \tau')=-\frac{1}{4\pi^2}\frac{1}{(\Delta \tau- i\eta)^2-(x-x')^2-(y-y')^2-(z-z')^2}~.
\end{eqnarray}
Now in the motion of circular Unruh quantum heat engine, we have
\begin{eqnarray}\label{trajectory1}
    t(\tau)=\gamma \tau,\,\,\, x(\tau)=R\cos{\left(\frac{v\gamma \tau}{R}\right)},\,\,\, y(\tau)=R\sin{\left(\frac{v\gamma \tau}{R}\right)},\,\,\, z(\tau)=0~.
\end{eqnarray}
Putting these expression in the Wightman function, we can write
\begin{eqnarray}\label{C_Wightman}
    \mathcal{G}(\tau, \tau')&=&-\frac{1}{4\pi^2}\frac{1}{(\gamma\Delta \tau- i\eta)^2-(R\cos{\frac{v\gamma \tau}{R}}-R\cos{\frac{v\gamma \tau'}{R}})^2-(R\sin{\frac{v\gamma \tau}{R}}-R\sin{\frac{v\gamma \tau'}{R}})^2}\nonumber\\
&=&-\frac{1}{4\pi^2}\frac{1}{(\gamma\Delta \tau- i\eta)^2-R^2(2-2\cos{\frac{v\gamma \tau}{R}}\cos{\frac{v\gamma \tau'}{R}}-2\sin{\frac{v\gamma \tau}{R}}\sin{\frac{v\gamma \tau'}{R}})}\nonumber\\
&=&-\frac{1}{4\pi^2}\frac{1}{(\gamma\Delta \tau- i\eta)^2-R^2(2-2\cos{\frac{v\gamma (\tau-\tau')}{R}})}\nonumber\\
&=&-\frac{1}{4\pi^2}\frac{1}{(\gamma\Delta \tau- i\eta)^2-4R^2\sin^2{\frac{v\gamma (\tau-\tau')}{2R}}}~.
\end{eqnarray}
 We take the ultra relativistic limit $\gamma>>1$  \cite{bell1983electrons}, making the $i\eta$ term insignificant. In this case we can see that only for $\tau-\tau'=\Delta \tau \xrightarrow{} 0$, we will get a contribution from this term while calculating the response function $\mathcal{F}$. For finite $\Delta \tau$, the first term in the denominator $\gamma^2\Delta \tau^2$ will dominate and make the whole term go to zero.

\noindent In this limit, we can expand the term $\sin^2\left[\frac{v\gamma (\tau-\tau')}{2R}\right]$ and write down the denominator of eq.\eqref{C_Wightman} in the following form.
\begin{eqnarray}
&=& \gamma^2\Delta\tau^2-4R^2\sin^2\left[\frac{v\gamma \Delta\tau}{2R}\right]=\gamma^2\Delta\tau^2-4R^2\left[\frac{v^2\gamma^2\Delta\tau^2}{4R^2}-\frac{v^4\gamma^4\Delta\tau^4}{48R^4}\right]\nonumber\\
&=&\gamma^2\Delta\tau^2-v^2\gamma^2\Delta\tau^2+\frac{v^4\gamma^4\Delta\tau^4}{12R^2}\nonumber\\
&=&\gamma^2\Delta\tau^2(1-v^2)+\frac{a^2\Delta\tau^4}{12},\hspace{5mm}(\text{where}\,\,\, a=\frac{v^2\gamma^2}{R}\xrightarrow{}\text{Acceleration})\nonumber\\
&=&\Delta\tau^2+\frac{a^2\Delta\tau^4}{12}~.
\end{eqnarray}
Thus, in the ultra-relativistic scenario the Wightmann function takes the form
\begin{eqnarray}
    \mathcal{G}^+(\tau, \tau')=-\frac{1}{4\pi^2}\frac{1}{\Delta\tau^2+\frac{a^2\Delta\tau^4}{12}}~.
\end{eqnarray}
Introducing the Lorentzian switching function \cite{arias2018unruh,mukherjee2022unruh}
\begin{eqnarray}
    \varpi_{\mathcal{T}}(\tau)=\frac{(\mathcal{T} / 2)^2}{\tau^2+(\mathcal{T} / 2)^2}
\end{eqnarray}
the response function of the detector can rewritten as
\begin{eqnarray}
    \mathcal{F}(\mathcal{E}, \mathcal{T})=\frac{1}{\hbar^2}\int_{-\infty}^\infty d\tau \int_{-\infty}^\infty d\tau'\left[\frac{(\mathcal{T} / 2)^2}{\tau^2+(\mathcal{T} / 2)^2}\frac{(\mathcal{T} / 2)^2}{\tau'^2+(\mathcal{T} / 2)^2}\right]\mathcal{G}^+(\tau, \tau'))e^{-\frac{i\mathcal{E} \Delta\tau}{\hbar}}~.
\end{eqnarray}
To compute this, we make the the variable transformation $(\tau, \tau' \xrightarrow[]{}m,n)$, where $m=\tau-\tau'$ and $n=\tau+\tau'$.\\

\noindent One of the advantages of the Lorentzian switching function is that it allows us to extend the integration to the complex plane and use the Cauchy residue theorem.

\noindent Following the calculation done in \cite{mukherjee2022unruh}, we can write
\begin{eqnarray}
    \mathcal{F}(\mathcal{T})&=&\frac{\pi \mathcal{T}^3}{4}\int_{-\infty}^{\infty}dm \frac{\mathcal{G}(m)}{m^2+\mathcal{T}^2}e^{-i\mathcal{E} m}\nonumber\\
&=&-\frac{ \mathcal{T}^3}{16\pi}\int_{-\infty}^{\infty}dm \frac{e^{-i\mathcal{E} m}}{(m^2+\mathcal{T}^2)(m^2+\frac{a^2m^4}{12})}\nonumber\\
&=&-\frac{ \mathcal{T}^3}{16\pi}\int_{-\infty}^{\infty}dm \frac{e^{-i\mathcal{E} m}}{(m^2+\mathcal{T}^2)m^2(1+\frac{a^2m^2}{12})}~.
\end{eqnarray}
We can compute this integral if we shift the pole at $m=0$ by $i\eta$ and take the limit $\eta\xrightarrow{}0$. Doing this we have
\begin{eqnarray}
    \mathcal{F}(\mathcal{T})&=&-\lim_{\eta\xrightarrow{}0}\frac{ \mathcal{T}^3}{16\pi}\int_{-\infty}^{\infty}dm \frac{e^{-i\mathcal{E} m}}{(m^2+\mathcal{T}^2)(m-i\eta)^2(1+\frac{a^2m^2}{12})}\nonumber\\
&=&-\lim_{\eta\xrightarrow{}0}\frac{ \mathcal{T}^3}{16\pi}\frac{12}{a^2}\int_{-\infty}^{\infty}dm \frac{e^{-i\mathcal{E} m}}{(m^2+\mathcal{T}^2)(m-i\eta)^2(m^2+\frac{12}{a^2})}~.
\end{eqnarray}
Due to the $e^{-i\mathcal{E} m}$ term, only the poles at the lower half of the imaginary plane will contribute to the contour integration. Thus the poles are at $m=-i\mathcal{T}$, $m=-i\frac{\sqrt{12}}{a}$~.
\vspace{0.7cm}
\begin{center}
\begin{tikzpicture}
    \draw[<->] (3,0) -- (11,0) node[right] {Re$[m]$};
    \draw[<->] (7,-3.5) -- (7,3) node[above] {Im$[m]$};
    
    \filldraw[red] (7,-2) circle (2pt) node[right] {$-i\mathcal{T}$};
    \filldraw[blue] (7,-1) circle (2pt) node[right] {$-i\frac{\sqrt{12}}{a}$};
    \draw[dashed] (4,0) arc[start angle=180, end angle=360, radius=3];
    \filldraw[red] (7,2.2) circle (2pt) node[right] {$i\mathcal{T}$};
    \filldraw[blue] (7,1.2) circle (2pt) node[right] {$i\frac{\sqrt{12}}{a}$};
    \filldraw[black] (7,0.2) circle (2pt) node[right] {$i\eta$};

\end{tikzpicture}
\end{center}
\vspace{0.7cm}

Residue from $m=-i\mathcal{T}$ reads
\begin{eqnarray}
R_1&=&-\lim_{\eta\xrightarrow{}0}\frac{ \mathcal{T}^3}{16\pi}\frac{12}{a^2}\lim_{z\xrightarrow{}-i\mathcal{T}}(z+i\mathcal{T}) \frac{e^{-i\mathcal{E} z}}{(z+i\mathcal{T})(z-i\mathcal{T})(z-i\eta)^2(z^2+\frac{12}{a^2})}\nonumber\\
&=&-\lim_{\eta\xrightarrow{}0}\frac{ \mathcal{T}^3}{16\pi}\frac{12}{a^2}\lim_{z\xrightarrow{}-i\mathcal{T}} \frac{e^{-i\mathcal{E} z}}{(z-i\mathcal{T})(z-i\eta)^2(z^2+\frac{12}{a^2})}\nonumber\\
 &=&-\frac{ \mathcal{T}^3}{16\pi}\frac{12}{a^2}\lim_{\eta\xrightarrow{}0} \frac{e^{-\mathcal{E} \mathcal{T}}}{(-2i\mathcal{T})(-i\mathcal{T}-i\eta)^2(-\mathcal{T}^2+\frac{12}{a^2})}\nonumber\\
&=&-\frac{ \mathcal{T}^3}{16\pi}\frac{12}{a^2} \frac{e^{-\mathcal{E} \mathcal{T}}}{(2i\mathcal{T}^3)(-\mathcal{T}^2+\frac{12}{a^2})}\nonumber\\
&=&-\frac{ 1}{16\pi}\frac{12}{a^2} \frac{e^{-\mathcal{E} \mathcal{T}}}{2i(-\mathcal{T}^2+\frac{12}{a^2})}~.
\end{eqnarray}

Residue from $m=-i\frac{\sqrt{12}}{a}$ reads
\begin{eqnarray}
R_2&=&-\lim_{\eta\xrightarrow{}0}\frac{ \mathcal{T}^3}{16\pi}\frac{12}{a^2}\lim_{z\xrightarrow{}-i\frac{\sqrt{12}}{a}}(z+i\frac{\sqrt{12}}{a}) \frac{e^{-i\epsilon z}}{(z^2+\mathcal{T}^2)(z-i\eta)^2(z+i\frac{\sqrt{12}}{a})(z-i\frac{\sqrt{12}}{a})}\nonumber\\
&=&-\lim_{\eta\xrightarrow{}0}\frac{ \mathcal{T}^3}{16\pi}\frac{12}{a^2}\lim_{z\xrightarrow{}-i\frac{\sqrt{12}}{a}} \frac{e^{-i\epsilon z}}{(z^2+\mathcal{T}^2)(z-i\eta)^2(z-i\frac{\sqrt{12}}{a})}\nonumber\\
&=&-\frac{ \mathcal{T}^3}{16\pi}\frac{12}{a^2}\lim_{\eta\xrightarrow{}0} \frac{e^{-\epsilon \frac{\sqrt{12}}{a}}}{(-\frac{12}{a^2}+\mathcal{T}^2)(-\frac{12}{a^2})(-2i\frac{\sqrt{12}}{a})}\nonumber\\
&=&-\frac{ \mathcal{T}^3}{16\pi} \frac{e^{-\epsilon \frac{\sqrt{12}}{a}}}{(-\frac{12}{a^2}+\mathcal{T}^2)2i\frac{\sqrt{12}}{a}}~.
\end{eqnarray}

Thus, from these residues we can write
\begin{eqnarray}\label{Fpos}
\mathcal{F}(\mathcal{E}, \mathcal{T})&=&-2\pi i(R_1+R_2)\nonumber\\
&=&-2\pi i(-\frac{ 1}{16\pi}\frac{12}{a^2} \frac{e^{-\mathcal{E} \mathcal{T}}}{2i(-\mathcal{T}^2+\frac{12}{a^2})}-\frac{ \mathcal{T}^3}{16\pi} \frac{e^{-\mathcal{E} \frac{\sqrt{12}}{a}}}{(-\frac{12}{a^2}+\mathcal{T}^2)2i\frac{\sqrt{12}}{a}})\nonumber\\
&=&\frac{ 1}{16}\left(\frac{12}{a^2} \frac{e^{-\mathcal{E} \mathcal{T}}}{(-\mathcal{T}^2+\frac{12}{a^2})}+\mathcal{T}^3\frac{e^{-\mathcal{E} \frac{\sqrt{12}}{a}}}{(-\frac{12}{a^2}+\mathcal{T}^2)\frac{\sqrt{12}}{a}}\right)~.
\end{eqnarray}

For $\mathcal{E}<0$, we have three poles, $m=i\mathcal{T}$, $m=i\eta$(second order) and $m=i\frac{\sqrt{12}}{a}$~.

\vspace{0.7cm}
\begin{center}
\begin{tikzpicture}
    
    \draw[<->] (3,0) -- (11,0) node[right] {Re$[m]$};
    \draw[<->] (7,-3) -- (7,3.5) node[above] {Im$[m]$};

    \filldraw[red] (7,2.2) circle (2pt) node[right] {$i\mathcal{T}$};
    \filldraw[blue] (7,1.2) circle (2pt) node[right] {$i\frac{\sqrt{12}}{a}$};
    \filldraw[black] (7,0.2) circle (2pt) node[right] {$i\eta$};
    \draw[dashed] (10,0) arc[start angle=0, end angle=180, radius=3];
     \filldraw[red] (7,-2) circle (2pt) node[right] {$-i\mathcal{T}$};
    \filldraw[blue] (7,-1) circle (2pt) node[right] {$-i\frac{\sqrt{12}}{a}$};
    
\end{tikzpicture}
\end{center}
\vspace{0.7cm}

Residue from $m=i\mathcal{T}$ reads
\begin{eqnarray}
R_3&=&-\lim_{\eta\xrightarrow{}0}\frac{ \mathcal{T}^3}{16\pi}\frac{12}{a^2}\lim_{z\xrightarrow{}i\mathcal{T}}(z-i\mathcal{T}) \frac{e^{i|\mathcal{E}| z}}{(z+i\mathcal{T})(z-i\mathcal{T})(z-i\eta)^2(z^2+\frac{12}{a^2})}\nonumber\\
&=&-\lim_{\eta\xrightarrow{}0}\frac{ \mathcal{T}^3}{16\pi}\frac{12}{a^2}\lim_{z\xrightarrow{}i\mathcal{T}} \frac{e^{i|\mathcal{E}| z}}{(z+i\mathcal{T})(z-i\eta)^2(z^2+\frac{12}{a^2})}\nonumber\\
&=&-\frac{ \mathcal{T}^3}{16\pi}\frac{12}{a^2}\lim_{\eta\xrightarrow{}0} \frac{e^{-|\mathcal{E}| \mathcal{T}}}{(2i\mathcal{T})(i\mathcal{T}-i\eta)^2(-\mathcal{T}^2+\frac{12}{a^2})}\nonumber\\
&=&\frac{ \mathcal{T}^3}{16\pi}\frac{12}{a^2} \frac{e^{-|\mathcal{E}| \mathcal{T}}}{(2i\mathcal{T}^3)(-\mathcal{T}^2+\frac{12}{a^2})}\nonumber\\
&=&\frac{ 1}{16\pi}\frac{12}{a^2} \frac{e^{-|\mathcal{E}| \mathcal{T}}}{2i(-\mathcal{T}^2+\frac{12}{a^2})}~.
\end{eqnarray}
\\
Residue from $m=i\frac{\sqrt{12}}{a}$ reads
\begin{eqnarray}
 R_4&=&-\lim_{\eta\xrightarrow{}0}\frac{ \mathcal{T}^3}{16\pi}\frac{12}{a^2}\lim_{z\xrightarrow{}i\frac{\sqrt{12}}{a}}(z-i\frac{\sqrt{12}}{a}) \frac{e^{i|\mathcal{E}| z}}{(z^2+\mathcal{T}^2)(z-i\eta)^2(z+i\frac{\sqrt{12}}{a})(z-i\frac{\sqrt{12}}{a})}\nonumber\\
&=&-\lim_{\eta\xrightarrow{}0}\frac{ \mathcal{T}^3}{16\pi}\frac{12}{a^2}\lim_{z\xrightarrow{}i\frac{\sqrt{12}}{a}} \frac{e^{i|\mathcal{E}| z}}{(z^2+\mathcal{T}^2)(z-i\eta)^2(z+i\frac{\sqrt{12}}{a})}\nonumber\\
&=&-\frac{ \mathcal{T}^3}{16\pi}\frac{12}{a^2}\lim_{\eta\xrightarrow{}0} \frac{e^{-|\mathcal{E}| \frac{\sqrt{12}}{a}}}{(-\frac{12}{a^2}+\mathcal{T}^2)(-\frac{12}{a^2})(2i\frac{\sqrt{12}}{a})}\nonumber\\
&=&\frac{ \mathcal{T}^3}{16\pi} \frac{e^{-|\mathcal{E}| \frac{\sqrt{12}}{a}}}{(-\frac{12}{a^2}+\mathcal{T}^2)2i\frac{\sqrt{12}}{a}}~.
\end{eqnarray}
\\
Residue from $m=i\eta$ reads
\begin{eqnarray}
    R_5&=&-\frac{ \mathcal{T}^3}{16\pi}\frac{12}{a^2}\frac{i|\mathcal{E}|a^2}{12T^2}\nonumber\\
    &=&-\frac{i|\mathcal{E}|T}{16\pi}~.
\end{eqnarray}
\\
Thus, from these residues we can write
\begin{eqnarray}\label{Fneg}
\mathcal{F}(-|\mathcal{E}|, \mathcal{T})&=&2\pi i(\frac{ 1}{16\pi}\frac{12}{a^2} \frac{e^{-|\mathcal{E}| \mathcal{T}}}{2i(-\mathcal{T}^2+\frac{12}{a^2})}+\frac{ \mathcal{T}^3}{16\pi} \frac{e^{-|\mathcal{E}| \frac{\sqrt{12}}{a}}}{(-\frac{12}{a^2}+\mathcal{T}^2)2i\frac{\sqrt{12}}{a}}-\frac{i|\mathcal{E}|\mathcal{T}}{16\pi})\nonumber\\
&=&\frac{1}{16}\left(\frac{12}{a^2} \frac{e^{-|\mathcal{E}| \mathcal{T}}}{(-\mathcal{T}^2+\frac{12}{a^2})}+ \mathcal{T}^3 \frac{e^{-|\mathcal{E}| \frac{\sqrt{12}}{a}}}{(-\frac{12}{a^2}+\mathcal{T}^2)\frac{\sqrt{12}}{a}}+2|\mathcal{E}|\mathcal{T}\right)~.
\end{eqnarray}
Therefore, using eq.(s)(\ref{Fpos}, \ref{Fneg}) into eq.\eqref{deltap}, we get
\begin{eqnarray}\label{deltapH}
    \frac{\delta p_{\mathcal{H}}}{\lambda^2}&=&\frac{ 1}{16}\Bigg[\frac{12}{a^2(-\mathcal{T}^2+\frac{12}{a^2})}\Big[(1-p)e^{-\mathcal{E} \mathcal{T}}-pe^{-|\mathcal{E}| \mathcal{T}}\Big]+\frac{\mathcal{T}^3}{(-\frac{12}{a^2}+\mathcal{T}^2)\frac{\sqrt{12}}{a}}\Big[(1-p)e^{-\mathcal{E} \frac{\sqrt{12}}{a}}-pe^{-\vert\mathcal{E}\vert \frac{\sqrt{12}}{a}}\Big]\Bigg.\nonumber\\
    &-&\Bigg.2p|\mathcal{E}|\mathcal{T}\Bigg]~.
\end{eqnarray}
\section{Analysis of the thermodynamics steps}\label{sec:analysis}
Now we will analyse the steps of the circular Unruh quantum Otto engine. We will also calculate the amount of heat exchanged between the detector and the two reservoirs and the resulting work extraction.

\subsection{Adiabatic expansion} In the first step, the initial density matrix of the detector remains unchanged $\rho_0=p\ketbra{e}{e} +(1-p)\ketbra{g}{g}$. The detector is moving with velocity $v$ along the path as indicated in the detector's cyclic motion.  However, over the course of this state, the energy difference between the two energy levels of the two-level detector is raised from $\mathcal{E}_1$ to $\mathcal{E}_2$.

\noindent Thus, we can write the time dependent Hamiltonian as
\begin{eqnarray}
    \mathcal{H}(t)=0\ketbra{g}{g}+\mathcal{E}(t)\ketbra{e}{e}~.
\end{eqnarray}

\noindent The nature of the increment of $\mathcal{E}_1$ to $\mathcal{E}_2$ will depend upon the function dependence of $\mathcal{E}(t)$ on t.

\noindent Now using the definition of average heat transfer, we can show
\begin{eqnarray}
    \langle Q_1\rangle=\int dt\,\mathbf{Tr}\left[\frac{d\rho_0}{dt}\mathcal{H}(t)\right]=0~.
\end{eqnarray}
The positive work done on the detector is given by
\begin{eqnarray}
    \langle W_1\rangle&=&\int dt\, \mathbf{Tr}\left[\rho_0\frac{d\mathcal{H}(t)}{dt}\right]   \nonumber\\
&=&\int dt\, \mathbf{Tr}\left[\rho_0\frac{d\mathcal{E}(t)}{dt}\ketbra{e}{e}\right]\nonumber\\
&=& p(\mathcal{E}_2-\mathcal{E}_1)~.
\end{eqnarray}

\subsection{Contact with the hot vacuum} In the second step, the detector is completing half rotation along the circumference of a circle with radius $R_1$. In this step the Hamiltonian of the system is constant $\mathcal{H}=\mathcal{E}_2\ketbra{e}{e}$ and the detector is interacting with the quantum field, moderated by the switching function. The detector's state $\rho$ changes due to the interaction with the field and at the end takes the form

\begin{eqnarray}
    \rho_{\mathcal{T}_1}=\begin{pmatrix}
p+\delta p_{\mathcal{T}_1} & 0 \\
0 & 1-p-\delta p_{\mathcal{T}_1}
\end{pmatrix}~.
\end{eqnarray}

\noindent As this is the hot reservoir, $\delta p_{\mathcal{T}_1}=\delta p_H$. In this step. The system absorbs heat from the vacuum, given by

\begin{eqnarray}\label{input_heat}
    \langle Q_2\rangle&=&\int dt\, \mathbf{Tr}\left[\frac{d\rho_0}{dt}\mathcal{H}(t)\right]\nonumber\\
    &=&\mathcal{E}_2\delta p_{\mathcal{T}_1}=\mathcal{E}_2\delta p_H~. 
\end{eqnarray}
\begin{eqnarray}
    \langle W_2\rangle=\int dt\, \mathbf{Tr}\left[\rho_0\frac{d\mathcal{H}(t)}{dt}\right]=0~.   
\end{eqnarray}

\subsection{Adiabatic Contraction} During this step the detector travels with velocity v in the opposite direction of step 1. During this the energy difference between the two levels are brought down to $\mathcal{E}_1$ from $\mathcal{E}_2$ and the density matrix of the detector remains unchanged. Similar to the adiabetic expansion, the heat transfer and the work can be calculated as 
\begin{eqnarray}
    \langle Q_3\rangle&=&\int dt\, \mathbf{Tr}\left[\frac{d\rho_0}{dt}\mathcal{H}(t)\right]=0  \\
    \langle W_3\rangle&=&-(p+\delta p_H)(\mathcal{E}_2-\mathcal{E}_1)~.
\end{eqnarray}
The negetive sign of $\langle W_3 \rangle $ indicates that the the work is done by the system.

\subsection{Contact with the cold vacuum} In the last step, the detector is completing half rotation along the circumference of a circle with radius $R_2$(Where $R_2<R_1$). In this step the Hamiltonian of the system is constant $\mathcal{H}=\mathcal{E}_1\ketbra{e}{e}$ and the detector is interacting with the quantum field, moderated by the switching function. The detector's state $\rho$ changes due to the interaction with the field and at the end takes the form
\begin{eqnarray}
    \rho_{\mathcal{T}_2}=\begin{pmatrix}
p+\delta p_{\mathcal{T}_1}+\delta p_{\mathcal{T}_2} & 0 \\
in0 & 1-p-\delta p_{\mathcal{T}_1}-\delta p_{\mathcal{T}_2}
\end{pmatrix}~.
\end{eqnarray}
As this is the cold reservoir,$ \delta p_{\mathcal{T}_1}=\delta p_C$. \\
The heat transfer and the work done is given by
\begin{eqnarray}
    \langle Q_4\rangle&=&\int dt\, \mathbf{Tr}\left[\frac{d\rho_0}{dt}\mathcal{H}(t)\right]=\mathcal{E}_1\delta p_C=-\mathcal{E}_1\delta p_H\\
    \langle W_4\rangle&=&0~.
\end{eqnarray}
The negative sign of $\langle Q_4\rangle$ is indicating that the system is dumping heat in the cold reservoir.

\subsection{Completing the cycle}
So far we have calculated and got the amount of heat exchanged
and work done in each step of the thermodynamical cycle. 

\noindent For the detector to return to its original state after completing the full cycle, we must impose the condition that $\delta p_H+\delta p_C=0$. Thus, the total amount of work done by the cycle is given by
\begin{eqnarray}
     \langle W_{total}\rangle&=&\langle W_1\rangle+\langle W_2\rangle+\langle W_3\rangle+\langle W_4\rangle\nonumber\\
    &=&-\delta p_H(\mathcal{E}_2-\mathcal{E}_1)~.\label{C_Work}
\end{eqnarray}

\noindent And the amount of heat transferred in the cycle is given by
\begin{eqnarray}
    \langle Q_{total}\rangle&=&\langle Q_1\rangle+\langle Q_2\rangle+\langle Q_3\rangle+\langle Q_4\rangle\nonumber\\
&=&\delta p_H(\mathcal{E}_2-\mathcal{E}_1)~.
\end{eqnarray}
From this we can see that
\begin{eqnarray}\label{engy_cons}
    \langle W_{total}\rangle+\langle Q_{total}\rangle=0~.
\end{eqnarray}
This confirms that the conservation of energy is maintained.
\section{Findings}\label{sec:result} In this section, we analyse our findings for the UQOE operating along a circular trajectory. We will analyze the transition probability and work extraction in the ultra-relativistic limit.
\subsection{Transition probability} It is of interest to us to analyse the transition probability during one of these circular motions. We analyse the change in transition probability $\delta p/\lambda^2$ with changing acceleration. We are specifically interested in three scenarios, $p=0$, $0<p<\frac{1}{2}$ and $\frac{1}{2}<p<1$.

\begin{figure}[h!]
  \centering
  \includegraphics[width=0.55\textwidth]{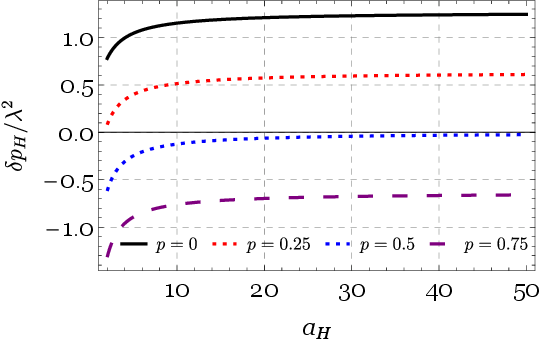}
  \caption{Behaviour of the transition probability with respect to the acceleration $a_H$ for various values of $p$ and $v=0.999$.}
  \label{fig:trans_prob}
\end{figure}

In Figure \eqref{fig:trans_prob} we have plotted $\delta p_H/\lambda^2$ for four different value of $p=0, 0.25, 0.5, 0.75$.  For the scenario $p=0$, we can say that the excited state is initially  empty and when the detector starts interacting with the vacuum during the circular motion, it is in its ground state. Now as we can see the transition probability of the detector is higher when the lower energy state is not highly populated. This is what we can expect from this scenario, as detectors with low population ground state will more likely to get excited to excited states. For the detectors with highly populated excited states($p>\frac{1}{2}$) it is not possible for them to have a positive transition rate $\delta p_H$ irrespective of the acceleration value as shown in the figure. One interesting case is when the value of $p$ is at the critical value of $\frac{1}{2}$, in this case the detector's ground state is half-filled, thus at finite values of the acceleration, the transition probability is negative and as the acceleration approaches infinity, it approaches to zero.

\begin{figure}[h!]
  \centering
  \includegraphics[width=0.55\textwidth]{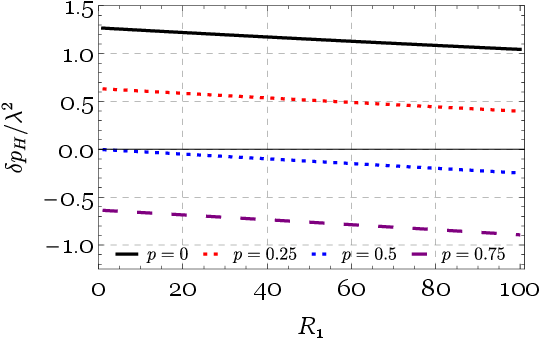}
  \caption{Behaviour of the transition probability with respect to the radius $R_1$ for various values of $p$ and $v=0.999$.}
  \label{fig:trans_prob1}
\end{figure}

We can also see the behavior of $\delta p_H/\lambda^2$ as a function of radius for different initial populations of the excited state. As we can see the acceleration $a$ is inversely related to the radius of the circular path. Thus, for the low acceleration values, the cycle can not produce positive values of $\delta p_H$. As the higher energy state of the detector is becoming less populated, it is becoming easier for the cycle to populate the upper state. As we can see from Figure \eqref{fig:trans_prob1} for the lowly populated states($p<\frac{1}{2}$) the transition probability is positive for the small values of radius (higher accelerations). 

\noindent However, for $p>\frac{1}{2}$, even for lower radius, the transition probability is less than zero, meaning the detector cannot absorb heat from the hot bath. Thus, from the graph, we can conclude that to extract heat we must have $0<p<\frac{1}{2}$.
\subsection{Work Extraction}
In this subsection, we calculate the work output of the UQOE along a circular motion in the ultra-relativistic scenario. Earlier we observed that the thermodynamic steps of the detector are unchanged by the effect of circular motion.

\noindent From our earlier analysis we know that the extracted work is given by
\begin{eqnarray}
    <W_{ext}>&=&\lambda^{2}\Delta\mathcal{E}\Big[(1-p) \mathcal{F}(\mathcal{E}, \mathcal{T})-p \mathcal{F}(-\mathcal{E}, \mathcal{T})\Big]\nonumber\\
&=&\lambda^{2}\Delta\mathcal{E}\left[(1-p) 
\left(\frac{ 1}{16}\frac{12}{a^2} \frac{e^{-\mathcal{E} \mathcal{T}}}{(-\mathcal{T}^2+\frac{12}{a^2})}+\frac{ \mathcal{T}^3}{16} \frac{e^{-\mathcal{E} \frac{\sqrt{12}}{a}}}{(-\frac{12}{a^2}+\mathcal{T}^2)\frac{\sqrt{12}}{a}}\right)-p \frac{1}{16}\left(\frac{12}{a^2} \frac{e^{-|\mathcal{E}| \mathcal{T}}}{(-\mathcal{T}^2+\frac{12}{a^2})}\right.\right.\nonumber\\
&+& \left.\left.\mathcal{T}^3 \frac{e^{-|\mathcal{E}| \frac{\sqrt{12}}{a}}}{(-\frac{12}{a^2}+\mathcal{T}^2)\frac{\sqrt{12}}{a}}+2|\mathcal{E}|\mathcal{T}\right)\right]~.
\end{eqnarray}
Here we have chosen $p$ arbitrarily. However, if we have to choose $p=p_{cyc}$ such that the detector returns to its original state satisfying the cyclicity condition, we must have 
\begin{eqnarray}
    \rho_{\mathcal{T}_2}=\rho_{\mathcal{T}_1}~.
\end{eqnarray}
Using the matrix form of $\rho_{\mathcal{T}_1}$ and $\rho_{\mathcal{T}_2}$, we get
\begin{eqnarray}
    \begin{pmatrix}
p+\delta p_{\mathcal{T}_1} & 0 \\
0 & 1-p-\delta p_{\mathcal{T}_1}
\end{pmatrix}=\begin{pmatrix}
p+\delta p_{\mathcal{T}_1}+\delta p_{\mathcal{T}_2} & 0 \\
0 & 1-p-\delta p_{\mathcal{T}_1}-\delta p_{\mathcal{T}_2}
\end{pmatrix}~.
\end{eqnarray}
This gives the condition for the cycle to be complete
\begin{eqnarray}
    \delta p_{\mathcal{T}_1}+\delta p_{\mathcal{T}_2}&=&0 \\
\Rightarrow \delta p_H+\delta p_C&=&0~.
\end{eqnarray}
Using this cyclicity condition, we can figure out the value of $p_{cyc}$. Using the expression for $\delta p$, and putting it in the cyclicity condition, we get
\begin{eqnarray}
    \lambda^2\Big[(1-p_{cyc})\mathcal{F}(\mathcal{E}_2,\mathcal{T}_2)-p_{cyc}\mathcal{F}(-\mathcal{E}_2,\mathcal{T}_2)\Big]+\lambda^2\Big[(1-p_{cyc})\mathcal{F}(\mathcal{E}_1,\mathcal{T}_1)-p_{cyc}\mathcal{F}(-\mathcal{E}_1,\mathcal{T}_1)\Big]=0~.
\end{eqnarray}
From the above equation, $p_{cyc}$ takes the form
\begin{eqnarray}
    p_{cyc}=\left[\frac{\mathcal{F}(\mathcal{E}_2,\mathcal{T}_2)+\mathcal{F}(\mathcal{E}_1,\mathcal{T}_1)}{\mathcal{F}(\mathcal{E}_2,\mathcal{T}_2)+\mathcal{F}(-\mathcal{E}_2,\mathcal{T}_2)+\mathcal{F}(\mathcal{E}_1,\mathcal{T}_1)+\mathcal{F}(-\mathcal{E}_1,\mathcal{T}_1)}\right]~.
\end{eqnarray}

\begin{figure}[h!]
  \centering
  \includegraphics[width=0.55\textwidth]{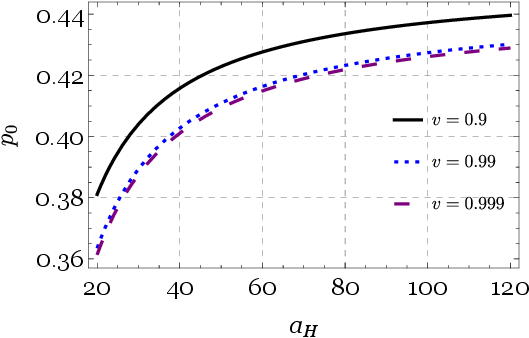}
  \caption{Behaviour of the critical probability with respect to the acceleration $a_H$ for various values of $v$ and $a_C=15$.}
  \label{fig:crit_prob}
\end{figure}

\noindent This recasted relation automatically ensures that the cyclicity condition is satisfied.
Now using the value of $p_{cyc}$ in the transition probability, it can be recast as
\begin{eqnarray}
    \delta \bar{p}_H=\lambda^2[(1-p_{cyc})\mathcal{F}(\mathcal{E},\mathcal{T})-p_{cyc}\mathcal{F}(-\mathcal{E},\mathcal{T})]~.
\end{eqnarray}
Using this equation, we can write the work extracted as 
\begin{eqnarray}
    W&\equiv& \langle W_{ext}\rangle=\delta \bar{p}_H (\mathcal{E}_2-\mathcal{E}_1)\nonumber\\
&=&\lambda^2\Big[(1-p_{cyc})\mathcal{F}(\mathcal{E},\mathcal{T})-p_{cyc}\mathcal{F}(-\mathcal{E},\mathcal{T})\Big](\mathcal{E}_2-\mathcal{E}_1)~.
\end{eqnarray}
Using this expression we have plotted the $\langle W_{ext}\rangle$ vs acceleration graph, where acceleration is given by $a=\frac{\gamma^2v^2}{R_1}$. From the description of the cycle we know that $\mathcal{T}$ and $R$ are related by $\mathcal{T}=\frac{\pi R}{\gamma v}$

\begin{figure}[h!]
  \centering
  \includegraphics[width=0.55\textwidth]{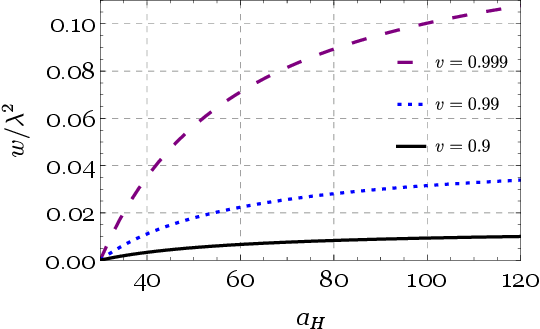}
  \caption{Behaviour of the output work with respect to the acceleration $a_H$ for various values of $v$ keeping $a_C=15$, $\Delta\mathcal{E}=1$.}
  \label{fig:work1}
\end{figure}

First let us look into the change in the critical probability $p_{cyc}$ with respect to the change in acceleration in the hot bath $a_H$ which is given in figure \eqref{fig:crit_prob}. As we can see with increasing acceleration the critical probability also increases for all the ultra-relativistic velocities that we have considered, $v=0.9, 0.99, 0.999$. However from the figure we can noting that as the detector approaches the speed of light, the critical probability at any given detector acceleration reduces systematically.

 Next when we try to analyze the behaviour of the extracted work $<W_{ext}>$,  we can see from the Figure \eqref{fig:work1} that the value of extracted work increases with acceleration and then reaches its asymptotic value for higher acceleration. In this case also we see a systematic change when the value of the detector velocities are considered. As the detector velocity approaches the speed of light, the work output of the circular UQOE increases significantly, indicating the the heat engine produces more work at extremely high velocities.
\subsection{Effect of energy gap in work extraction}
One of the most interesting results that we found in our analysis is that the amount of extracted work at a particular accelerations (or radius)
changes non-linearly with the energy level difference in the detector. 
\begin{figure}[h!]
  \centering
  \includegraphics[width=0.55\textwidth]{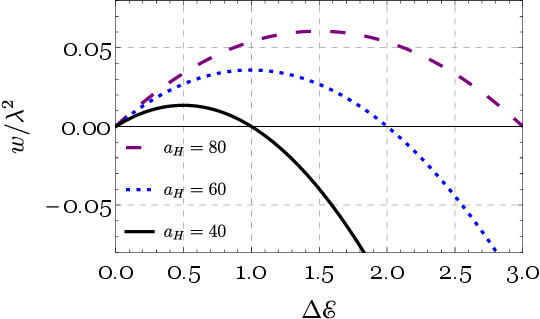}
  \caption{Behaviour of the output work with respect to the energy gap $\Delta\mathcal{E}$ for various values of $a_H$ keeping $a_C=20$, $v=0.999$.}
  \label{fig:work2}
\end{figure}

\noindent Note that the work output reaches a maximum value at a certain energy level difference. Value of this energy gap difference is dependent on the chosen acceleration acceleration of the two circular paths and thus are dependent on the two values of radius.

The possible explanation can be as the work done in the first adiabatic phase to change the energy gap of the detector from $\mathcal{E}_1$ to $\mathcal{E}_2$ ($\mathcal{E}_2>\mathcal{E}_1$) becomes larger with the increase in the value of $\mathcal{E}_2-\mathcal{E}_1\equiv \Delta\mathcal{E}$, the $\Delta\mathcal{E}$ part of the extracted work expression eq.\eqref{C_Work} increases. However,with this increment in the energy gap, when the detector experiences the hot bath, it becomes much more difficult for it to get excited to its excited state due to the heightened energy barrier between its two energy states. Thus, in eq. \eqref{C_Work} the $\delta p_H$ decrease with this energy gap increase. These two competing effects give us the point where the extracted work reaches a maximum, and its acceleration dependence arises from the fact that at different accelerations the detector experiences different temperatures in the hot bath, which dictates the detector's excitation probabilities.
\subsection{Efficiency}
In any thermal machine, whether classical or quantum, efficiency is a fundamental quantity of interest. In this model, the net work done by the engine and the total heat exchanged between the qubit and the quantum vacuum along its circularly accelerated motion are given by eqs. (\ref{input_heat}, \ref{C_Work}). Our analysis confirms that these quantities satisfy the principle of energy conservation, as expressed in Eq. \eqref{engy_cons}.  

From the expressions in eqs. (\ref{input_heat}, \ref{C_Work}), the efficiency of the heat engine can be determined. Since external stimulation is applied to modulate the energy gap of the qubit levels during the adiabatic expansion and contraction phases, as discussed in Section \ref{sec:analysis}, the work performed by the qubit is given by \(\langle W_{ext} \rangle = -\langle W \rangle\). Consequently, the efficiency is expressed as  

\begin{eqnarray}\label{c_eff}
    \eta = \frac{\langle W_{ext} \rangle}{Q_2} = 1 - \frac{\mathcal{E}_1}{\mathcal{E}_2}~.
\end{eqnarray}  

\noindent Thus, eq. \eqref{c_eff} demonstrates that the efficiency in our model is independent of the specific implementation of the heat engine. Instead, it solely depends on the ratio of the energy gaps between the two levels of the qubit and exhibits a form consistent with previous studies \cite{arias2018unruh,gray2018scalar,xu2020unruh,mukherjee2022unruh}.  

\section{Conclusion}\label{sec:conclu}  
In this study, we propose a novel model of the relativistic quantum Otto engine using the notion of the circular Unruh effect \cite{bell1983electrons,bell1987unruh}. As experimentally constructing and utilising a linearly accelerating Unruh-Dewitt detector in the Unruh Quantum otto Engine is not practical for several reasons, and was a long standing problem as we discussed earlier. In this article, we have introduced a relativistic quantum thermal machine based on the circular Unruh effect to circumvent these issues. 

In this model, we consider a qubit undergoing a closed one-dimensional trajectory while interacting with a massless quantum scalar field in two distinct regions. In these regions, the qubit experiences accelerations of different magnitudes, enabling the quantum field to act as an effective thermal reservoir. To incorporate the effects of the circular Unruh effect, we assume that the qubit follows circular motion along two distinct arcs with radii \( R_1 \) and \( R_2 \) while interacting with the massless quantum scalar field. As a result of this circular motion, the quantum vacuum effectively serves as a thermal reservoir, with the radii \( R_1 \) and \( R_2 \) corresponding to the hot and cold reservoirs of the engine, respectively. 

It has been previously observed that the correlation function of scalar fields, commonly referred to as the Wightman function, is highly dependent on the specific model considered \cite{rizzuto2007casimir, mukherjee2022unruh}. Since the response function of the qubit is directly influenced by this correlation function \cite{takagi1986vacuum, hummer2016renormalized, louko2016unruh, sachs2017entanglement}, the introduction of circular motion in the qubit's trajectory will lead to a corresponding modification in its response. Consequently, the transition probability will also be altered, capturing the effects of the qubit’s circular motion.

In this study, we have calculated the heat and work for each step of the cycle and analysed the engine's characteristics. We have used two circular paths in this model to make it experimentally viable, as this reduces the amount of space required to achieve the required acceleration while maintaining the cyclic nature of the engine. Then, using  quasi-static processes and perturbation theory, we are able to establish the conditions over the initial excitation probability $p_{cyc}$ and detector accelerations $a_H$ and $a_C$. We have also calculated the efficiency of the circular UQOE.

Our analysis reveals several significant findings. We observe that the transition probability of the qubit increases with its acceleration and eventually saturates in the high-acceleration regime. Notably, the transition probability remains positive for excitation probabilities $p=0$ and $p=1/2$. Furthermore, the transition probability decreases with an increasing radius of the arc in the hot bath. Since, in circular motion, the radius is inversely proportional to the centripetal acceleration, this behavior is expected. Additionally, our study shows that the critical excitation probability increases with the acceleration in the hot bath. As the qubit velocity approaches the speed of light, the critical excitation probability attains its lowest value, suggesting that the engine can operate efficiently even at low initial excitation probabilities.

Regarding energy extraction, we find that the output work increases with the acceleration of the hot bath while keeping the cold bath acceleration fixed. Moreover, the output work reaches its maximum at the highest qubit velocity. Our analysis further reveals the existence of an optimal energy gap between the detector's energy states that maximizes energy extraction. This suggests that, for specific detector velocities and accelerations, choosing appropriate atomic (qubit) systems with tailored energy gaps can significantly enhance energy extraction in experimental setups.

 The study of circular UQOE is also important because of its experimental value. Recently, quite a lot of advancement has been made in the  hyper-fast rotation in an optically levitated nanoparticle
 system \cite{GHz2018Reimann, OpticalLevitate2018Ahn} and superconducting qubits \cite{Relativistic2015qubits}. Building one of these experimental setups in the future, we can use the high rotational velocities achieved in these nanoparticles to set up the circular UQOE described in this paper. One of the possible ways to achieve it is to  attach the atom to a hyper-fast rotating nanoparticle \cite{lochan2020detecting}.

Our approach paves the way for several promising avenues of future research. While this study focuses on scalar quantum fields, further investigations could extend to alternative heat engine models incorporating diverse working substances and different quantum field theories, particularly electromagnetic fields. Additionally, exploring atoms (detectors) with more than two energy levels \cite{Varinder2020optimal} or degenerate energy levels \cite{xu2020unruh} could offer deeper insights, especially in the context of potential experimental realizations. Another compelling direction involves analyzing heat engines operating in varied gravitational backgrounds, such as curved spacetime \cite{kollas2024gravitational}, including black hole environments. Advancing research in this area could further illuminate the intricate connections between thermodynamics and gravitational theories.

\section*{Acknowledgement}
\noindent RPS and AM would like to thank S. N. Bose National Centre for Basic Sciences for providing the financial support.

\bibliographystyle{hephys}
\bibliography{Master_Reference}
\end{document}